# Mechanical conversion of low-affinity Integration Host Factor binding sites into high-affinity sites

Merek Siu*[1], Hari Shroff *[1], Jake Siegel[1], Ann McEvoy[1],
David Sivak[1], Ann Maris[2], Andrew Spakowitz[3], Jan Liphardt[1,2,4]

1. Biophysics Graduate Group, UC Berkeley, Berkeley CA.
2. Physical Biosciences Division, Lawrence Berkeley National Laboratory, Berkeley CA.
3. Dept. of Chemical Engineering, Stanford University, Stanford, CA.
4. Dept. of Physics, UC Berkeley, Berkeley, CA.
*These authors contributed equally to this work.

H. Shroff's current address is Janelia Farm Research Campus, Howard Hughes Medical Institute

Corresponding Author:
Jan Liphardt. Address: Physics Department, University of California, Berkeley, Berkeley, CA
94720. Tel: (510) 666-2784, Fax: (510) 643-8497, liphardt@physics.berkeley.edu

**Abstract**

Although DNA is often bent *in vivo*, it is unclear how DNA-bending forces modulate DNA-protein binding affinity. Here, we report how a range of DNA-bending forces modulates the binding of the Integration Host Factor (IHF) protein to various DNAs. Using solution fluorimetry and electrophoretic mobility shift assays, we measured the affinity of IHF for DNAs with different bending forces and sequence mutations. Bending force was adjusted by varying the fraction of double-stranded DNA in a circular substrate, or by changing the overall size of the circle (1). DNA constructs contained a pair of Förster Resonance Energy Transfer dyes that served as probes for affinity assays, and read out bending forces measured by optical force sensors (2). Small bending forces significantly increased binding affinity; this effect saturated beyond ~3 pN. Surprisingly, when DNA sequences that bound IHF only weakly were mechanically bent by circularization, they bound IHF more tightly than the linear "high-affinity" binding sequence. These findings demonstrate that small bending forces can greatly augment binding at sites that deviate from a protein's consensus binding sequence. Since cellular DNA is subject to mechanical deformation and condensation, affinities of architectural proteins determined *in vitro* using short linear DNAs may not reflect *in vivo* affinities.

## Introduction

DNA bending and looping play important roles in transcriptional regulation (3, 4), recombination (5), and chromosome compaction (6). The manner in which proteins induce sharp bends in their DNA substrates, and the converse problem of how DNA conformation affects protein binding affinity, have been extensively studied (7-16). Among other things, it is now understood that "pre-bending" DNA substrates can facilitate the formation of certain protein-DNA complexes. The increased affinity for pre-bent DNA has been demonstrated for catabolite activator protein (CAP) (15), TATA binding protein (TBP) (16), and Integration Host Factor (IHF) (14).

The next step in the field is to move beyond binary categories of "bent" or "unbent" DNA, and to quantify how protein-DNA affinities are modulated by a variety of forces. For instance, inside the cell, DNA compaction establishes a background DNA bending force that is continually changed by DNA-manipulating molecular machines. These forces may modulate protein-DNA affinity in ways that are difficult to gauge with conventional protein-DNA affinity assays. It is also important to determine whether sequence mutations that normally disrupt binding can be compensated by forces that bend the DNA.

Our model system is IHF, a 22kDa heterodimeric protein that site-specifically binds and sharply bends DNA (11, 17). IHF participates in site-specific recombination, transcriptional regulation, and chromosome compaction (17-24). The crystal structure of the IHF-DNA complex reveals that in it, the DNA is bent by 160 degrees into a U-turn (11); this bending has been corroborated by other techniques (13, 25, 26). The sites where IHF binds and sharply bends DNA are largely dictated by an indirect readout mechanism (27), which discriminates among DNA sequences via sequence-dependent architectural features including conformation (e.g. twist) and flexibility (28). Both the dramatic DNA deformation induced by IHF, and the importance of sequence-induced DNA structure and flexibility on binding site recognition (29), suggest IHF as a model system for examining the interplay between bending forces and the effects of DNA sequence.

We measured the affinity of IHF for substrates subject to different bending forces (Fig. 1) using a variety of biophysical techniques, including FRET, electrophoretic mobility shift assays (EMSA), and optical force sensors. A bending force, internal to the DNA construct, was changed by varying substrate geometry. Annealing different lengths of complementary DNA to single-stranded DNA loops yielded substrates with a variety of internal bending forces. The resulting circular constructs were hybrids of stiff double-stranded DNA and comparatively flexible single-stranded DNA. The loops also contained FRET donor and acceptor fluorophores that served two functions: as reporters for binding affinity assays, and as parts of an optical force sensor that reports the force internal to circular constructs (2). Whereas previous studies on IHF binding to pre-bent DNA used fully double-stranded mini-circles that were over 145 bp (13-16), we used much smaller constructs constrained to 57 nucleotides (nt) to obtain tightly bent substrates with adjustable and measurable internal forces.

Consistent with previous studies on CAP and TBP, we found that IHF binds strained circularized substrates much tighter than otherwise equivalent linear substrates. Small bending forces significantly boosted binding affinity. The effect was not linear and appeared to saturate at about three picoNewtons (pN): forces beyond 3 pN did not further increase binding affinity. Surprisingly, mechanical forces could be used to convert a mutant "low-affinity" binding site into a high affinity binding site, which bound IHF more tightly than the "high-affinity" consensus binding site for IHF. These findings have implications for the identification of relevant protein binding sites in genomes (30), could clarify how regulatory signals are integrated at transcription initiation sites (3), and may enable controlled mechanical tuning of the affinity of protein-DNA interactions.

**Results and Discussion**

Investigating the relationship among DNA bending force, DNA sequence, and protein-DNA affinity required synthesis of DNA constructs with varying internal bending forces (hereafter referred to as internal force). To that end, unstressed linear and stressed circular DNAs were prepared (Fig. 1). Circular substrates were made by ligating ssDNA and then annealing complementary DNAs (Fig. 1C). The internal force in a loop was controlled by either changing the fraction of the loop that was double-stranded or by changing the loop length. Since dsDNA is significantly stiffer than ssDNA (the persistence length of dsDNA is ~50 nm but the persistence length of ssDNA is ~1 nm) (31, 32), changing the ratio of dsDNA:ssDNA modulates the internal force.

Since the mechanics of small DNA circles are incompletely understood (although this is improving rapidly (1, 33-35)) it is not possible to accurately predict their internal forces. Hence, we measured them experimentally by integrating force-sensors that report force via optical changes. The force sensor consists of an entropic spring (a piece of ssDNA) with a pair of FRET dyes, such that the extension of the spring (and thus the force acting on the spring) can be tracked via FRET (2). At low forces, the entropic elasticity of the ssDNA brings the donor and acceptor dyes close together, resulting in high FRET efficiency. At high forces, the ssDNA is stretched, and FRET efficiency decreases.

The FRET vs. force curve of our force sensor could not be calculated due to the complicated effects of potential basepairing on its extensibility. Hence, we experimentally calibrated the sensor using single-molecule techniques (2). Known forces were applied to a variety of ssDNA springs with magnetic tweezers and the FRET efficiency was simultaneously monitored using single-molecule total internal reflection fluorescence. Using these optical force sensors, we previously showed that 101mer loops have internal forces around 1-3 pN, whereas shorter 57mer loops have internal forces between ~5-≥12 pN at low ionic strengths (1). In the present study, we use a force sensor with Cy3 and Cy5 dyes separated by ten single-stranded bases, which reports forces up to 20 pN (previously designated as FS10 (2)).

We used two complementary methods to determine the binding affinity of IHF to DNA substrates with varying internal forces. The first approach was a solution FRET assay that takes advantage of the FRET dyes that are part of the optical force sensor. Solution FRET has previously been used to examine the binding of IHF to linear substrates (26, 36-38). Here, we extend the technique to prebent circular DNA substrates. This assay yields true equilibrium measurements that do not suffer from dissociation that might occur during electrophoresis. Electrophoretic mobility shift assays, however, have the advantage of being able to detect binding stoichiometry and non-specific binding (39, 40); EMSA has been previously used with IHF (41-43). Since IHF can recognize DNA in both specific and non-specific modes (17, 42), EMSA was essential to confirm that we were probing specific interactions and complexes, as determined by their characteristic gel shift mobility. EMSA also permitted clean comparisons of linear and circular constructs whose sole difference is the internal force associated with their geometry. Both solution FRET and EMSA were performed in high salt buffers (220mM NaCl, pH 7.4) to minimize non-specific binding of IHF to DNA (44).

*IHF binds circular constructs with more than ten-fold higher affinity than linear constructs*

Circles produced by ligating ssDNAs were purified by gel electrophoresis. Ligation products were resistant to digestion by exonuclease 1, confirming their circular geometry. Annealing complementary DNA yielded a double-stranded region containing the well-characterized, high affinity λ H' binding site for IHF (11). The nomenclature of the constructs is as follows: the first number denotes the overall length of the DNA in nucleotides, "L" denotes linear, "C" denotes circular, and the second number denotes the length of the double-stranded region. Thus, the 57C35 substrate consists of a circular 57 nucleotide (nt) ssDNA to which a 35 nt complement has been annealed.

We first compared IHF binding to linear (43L41 - Fig. 1A or 57L35 - Fig. 1B) and circular (57C35 - Fig. 1C) DNA constructs with solution FRET (Fig. 2) and EMSA (Fig. 3). Solution FRET assays (Fig. 2) were performed by titrating concentrated IHF into a solution of dually-labeled DNA. For both linear (Fig. 1A) and circular (Fig. 1C) DNA constructs, the dramatic bend induced by IHF brings the donor and acceptor dyes together resulting in greater FRET as evidenced by reduced Cy3 emission (567 nm peak) and increased Cy5 emission (670 nm peak) (Fig. 2A&B).

Taking advantage of the different degrees of FRET between the free and bound state, we used a FRET ratio to indicate the fraction of bound DNA. The FRET ratio is the ratio of acceptor emission resulting from FRET (Cy3 excitation – Fig 2A&B), to acceptor emission when directly excited (Cy5 excitation – Fig. 2A&B Inset) (see also Methods). This measure accounts for dye bleaching and DNA dilution as the titration progresses.

Fitting the fraction bound (obtained by measuring the FRET ratio) to a one-site saturation binding model yields the dissociation constant ($K_D$) (see Methods). For the unstressed 43L41 substrate, the $K_D$ was $9.3 \pm 0.2$ nM (Table 1), on the same order as to the previously reported $K_D$ of 1.9 nM for a fully double-stranded 30mer containing the H' binding site (41). The slightly weaker binding affinity we observe is consistent with the higher, more stringent salt concentrations used in these experiments (220mM NaCl here versus 60 mM KCl in (41)). Control titrations of equivalent volumes of buffer showed no increase in FRET (Fig. 2C&D open circles).

When the DNA was strained by circularization (57C35, Fig. 1C), significantly less IHF was required to reach saturation binding (note the different scale on the x-axis in Fig. 2D vs. Fig. 2C). The dissociation constant for 57C35 decreased to $0.49 \pm 0.01$ nM. Hence, circularizing the DNA substrate increased IHF affinity almost 20-fold (Table 1). For comparison, a five-fold increase in affinity was observed for binding to 163 bp fully double-stranded mini-circles versus linear DNAs (14).

Similar results were obtained with an orthogonal method of measuring binding affinity - the electrophoretic mobility shift assay (EMSA). Figure 3 shows representative gels for linear, 57L35 (Fig. 3A), and strained circular 57C35 (Fig. 3B) constructs. For each gel, a constant concentration of IHF is titrated with increasing amounts of dye-labeled DNA. For both linear and circular constructs, DNA bound by IHF has a retarded mobility relative to free DNA.

We observed an increased affinity for the strained circular 57C35 (0.2 ± 0.1 nM) when compared to unstressed linear 57L35 (6.8 ± 2.3 nM). The similarity in dissociation constants obtained by solution FRET and EMSA methods is encouraging, especially given the different titration schemes (titrating DNA vs. titrating IHF) used in the different assays.

Comparison between the assays also serves as a control to check for binding to the ssDNA region between the FRET pair. The similarity in measured binding affinity for the 43L41 construct (solution FRET) and the 57L35 (EMSA) suggest that the single-stranded tail present with 57L35 does not noticeably affect IHF binding under these conditions. This was further confirmed with saturation binding assays on a completely single-stranded linear substrate, which revealed no evidence of IHF binding to the ssDNA region common to the 57mer constructs (Fig. S1). Finally, there was also no evidence for non-specific binding, dissociation during electrophoresis, and bound complexes other than the expected 1:1 IHF:DNA stoichiometry. These observations are important for a straightforward interpretation of the solution FRET assays.

The EMSA assays used linear constructs that were as similar as possible to the circular constructs. The 57L35 construct (Fig. 1B) is exactly the same as the 57C35 (Fig. 1C) construct except for the single phosphodiester bond that distinguishes their geometry. We did not use the 57L35 construct for solution FRET studies, since both dyes are on the same side of the IHF binding site (Fig. 1B); IHF binding and bending does not bring the dyes closer together with a resulting FRET change.

For the EMSA assays (but not the solution FRET studies), we note that the concentrations of IHF and DNA tend towards the "stoichiometric regime." Our experimental conditions were chosen to ensure sufficient fluorescent signal for visualization. Binding affinities can still be measured under these conditions provided that ligand depletion is taken into account (45) - the equilibrium is simply shifted. Importantly, we obtained similar absolute dissociation constants as well as identical trends between DNA constructs with the solution FRET and EMSA assays (Table 1).

*Complement lengths resulting in internal forces >3 pN do not significantly affect IHF binding affinity*

Having confirmed that IHF binds circularized DNA more tightly than linear DNA, we set out to determine if the affinity could be controlled by modifying the internal force of the DNA substrate. To vary the internal force, we adjusted the amount of double-stranded DNA in a given loop and the total loop length. We first investigated three additional ds:ssDNA ratios by annealing complements of 30, 39, and 43 nts to the 57mer loop. The minimum complement length, 30 nt, is determined by the footprint of IHF (46), and the maximum complement length of 43 nt is limited by the force sensor region, which must remain single-stranded to utilize the previous force calibrations (2). Annealing these selected complements yields constructs with internal forces of ~5-≥12 pN at low ionic strength (1).

Dissociation constants for 57C30, 57C39 and 57C43 were obtained by solution FRET titrations and EMSA (Fig. 2D and Table 1). EMSA was also performed for the equivalent linear constructs (57L30, 57L39 and 57L43). IHF affinity for DNA increased by at least an order of magnitude upon circularization, from 5-13 nM to 0.2-0.9 nM (ranges include both solution FRET and EMSA values, Table 1). Thus, across all complement lengths, IHF binds small strained circles with greater affinity than linear DNAs of identical sequence.

We also investigated IHF binding affinity to less strained 101mer loops with EMSA. 101C87 showed a roughly 10-fold increase in affinity ($K_D$ = 5.1 ± 1.7 nM for linear, and 0.5 ± 0.2 nM for circular), similar to the increase seen with the 57mer (Table 1). However, the 101C77 construct exhibited only a 1.5-fold increase in affinity, the smallest difference seen ($K_D$ = 1.5 ± 0.3 nM for linear to 0.9 ± 0.2 nM for circular). Affinity changes with smaller complements could not be measured because the electrophoretic mobility shifts were too small to be able to make reliable binding affinity measurements (Fig. S3). Likewise FRET changes upon binding of IHF were too small to obtain reliable affinity measurements.

There was no clear effect on binding affinity when either the dsDNA:ssDNA ratio or loop length was varied (Fig. 2D and Table 1). For the unstressed linear constructs, this agrees with a previous study which found similar affinities for linear 30 bp and 179 bp duplexes (41). The lack of a clear trend for the circularized constructs was surprising since circularized constructs with different complement lengths and loop sizes would be expected to have different internal forces and curvatures. To understand the null result, the lack of a clear affinity trend, we used the integrated optical force sensors to measure the forces internal to the loops.

FRET from the optical force sensors provides an estimate of the internal force in a circular DNA substrate prior to IHF binding. Different dsDNA:ssDNA ratios yield different FRET efficiencies (Fig. S2, (1)). These efficiencies are true FRET efficiencies (as opposed to the FRET ratio used in the solution FRET assays), where the degree of dye labeling and the polarization, extinction coefficients, and quantum yields of the dyes have been taken into account (1, 47).

To estimate internal force, we converted FRET efficiency to force via a previously determined calibration curve (2). A caveat in interpreting FRET efficiencies in terms of forces is a difference in buffer conditions: EMSA was performed in a high salt buffer to ensure that we were observing IHF binding in its specific binding mode whereas the sensor calibration experiments were done in a comparatively low salt buffer (2). Both dsDNA (48, 49) and ssDNA (32) exhibit a decreased persistence length under high salt conditions, because increased shielding of the phosphate backbone charges at higher ionic strength permits sharper bending. A decreased high-salt persistence length for both dsDNA and ssDNA will result in the dyes being closer together on average and result in a higher measured FRET value. To identify the forces in high salt buffer with greatest accuracy, a complete single molecule recalibration of the force sensor in high salt buffer would be required. Instead, we estimate the forces by using the FRET values measured in low salt buffer and the original force sensor calibration curve. Metropolis Monte Carlo simulations revealed negligible force changes to within the precision of the simulations (± ~1pN) when accounting for the expected persistence length changes of double and single-stranded DNA due to the different salt conditions (see Supplementary Materials and Table S1).

Using the previously measured calibration curve, we converted the relationship between affinity (from EMSA measurements) and FRET efficiency (Fig. 4A) to affinity versus internal force (Fig. 4B). Linear constructs are not plotted in Fig. 4A since their FRET efficiencies do not relate to internal force. Circularizing the constructs, yielding ~2-3 pN internal forces (101mers), was sufficient to increase the affinity 1.5-fold. Internal forces greater than ~3 pN (observed in the 57mer constructs) resulted in greater increases in affinity (~10-fold). Further increases in internal force (beyond 3pN) exhibited no significant effect (Fig. 4B). Note that the FRET efficiency measured for 57C30 maps to a force above the maximum force explored in the empirical calibration (2).

The significant increases in affinity observed upon circularization are consistent with the notion that internal force bends the DNA and therefore reduces the free energy of bending that must be expended by IHF. The relative insensitivity of IHF binding to forces greater than a few pN was unexpected. However, without high-resolution structural data of the final bound states of the various constructs, it is not particularly worthwhile to speculate about the mechanochemical basis for the effect's saturation at ~3-5 pN. It is possible that besides modulating the internal force within the loop, changing the complement length may affect substrate geometry in ways that could perturb IHF binding affinity. Since IHF is a minor-groove binding protein, the accessibility of the minor groove in the different constructs is an important consideration. Indeed, the correspondence between the direction of the bend induced by internal force and the final protein-DNA structure is important (15, 16). The relative position of where the ssDNA enters and leaves the dsDNA helix, which is related to the helical repeat of DNA, will affect the torsional strain of the construct. Since the loops are not rotationally constrained, relaxing to the construct's lowest torsional energy configuration will affect both the accessibility of the minor groove and the direction of pre-bend in a complement length dependent manner.

*Affinity reductions due to mutations can be rescued by internal force*

Last, we explored the sensitivity of internal force induced changes to affinity to another parameter: DNA sequence mutations. We studied two mutant sites (57M1 and 57M2), with significantly reduced binding affinities for IHF (41). Both mutants replace an AT basepair in the IHF consensus binding patches with a GC basepair. By modeling the IHF crystal structure complex (PDB ID 1IHF), the 57M1 mutant was predicted to disrupt the hydrogen bonding of an arginine inserted into the minor groove. Similarly, the 57M2 mutant was predicted to disrupt a proline intercalation site. Mutant circular constructs were variants on the 57C35 construct.

A comparison of solution FRET binding curves for unstressed wildtype (43L41), versus mutant M1 (43M1L41) constructs demonstrates the significantly reduced binding affinity of IHF to the M1 linear mutant (Fig. 5A) (43M1L41: 156 ± 6 nM, 43L41: 9.3 ± 0.2 nM). Our measured dissociation constant for the M1 linear construct is slightly higher than the previously reported value of 81 nM (41) (again consistent with the higher salt conditions used in this assay). In contrast, the solution FRET binding curves for the wildtype 57C35 circles and the mutant M1 circles (57M1C35) are similar (Fig. 5B) reflecting their similar dissociation constants (57M1C35: 0.89 ± 0.06 nM, 57C35: 0.49 ± 0.01 nM, Table 2). Internal force thus results in almost a 200-fold increase in affinity.

This demonstration of mechanical rescue of IHF binding affinity to mutant DNA sequences is corroborated by EMSA experiments. The significantly lower affinity of linear mutant constructs required the use of nonhomologous competition assays (Fig. S4) (2). We found that 57M1L binds IHF several-fold worse than the wildtype λ H' site (57L35) with an affinity of 46 ± 6 nM. Upon circularization, the $K_D$ of the 57M1L35 mutant dropped to 0.6 ± 0.4 nM, indicating a ~80-fold increase in affinity (Table 2). Remarkably, both solution FRET and EMSA binding assays show IHF bound the circularized "low-affinity" mutant site (57M1C35) about 10-fold *tighter* than a linear "high-affinity" consensus site (57L35 or 43L41).

The affinity of the 57M2L mutant was too low to measure accurately by competition EMSA or solution FRET. Instead, we use the previously measured value of 690 nM (which used radioactivity - a method incompatible with the force sensors) (41). This dissociation constant is likely a lower bound in our assay conditions, since the M2 binding affinity in our high salt buffer should be weaker than in the previous study's lower salt buffer. As with the M1 mutant, IHF bound the circularized 57M2C35 substrate tighter than the linear wildtype "high-affinity" binding site (43L41 or 57L35), which is remarkable since the affinity of IHF for the unstressed linear 57M2L35 site was two orders of magnitude *lower* than for the wild-type site (Fig. 5B inset, Table 2). The affinity increases observed upon circularizing the mutant constructs were the largest seen in this study. This raises the possibility that some DNA sequences conventionally believed to be deficient for IHF binding (as judged e.g. by gel retardation on short linear DNAs), may actually bind IHF more tightly *in vivo* than the supposedly high-affinity sites.

### Conclusions

These studies highlight the interplay between DNA substrate internal force, specificity of DNA sequence and protein-DNA binding affinity. The impact of internal force is most strikingly demonstrated by the rescue of binding affinity in circularized constructs mutated away from the wildtype sequence. In this light, it may be essential to account for DNA architecture when determining relevant protein binding sites *in vivo*. Unstressed DNA fragments or plasmids which have been traditionally used to map binding sites tend to be linear on the length scale of the binding site. For architectural proteins like IHF, mutational studies that use locally linear DNA substrates to highlight the importance of specific bases for protein site determination may be less relevant *in vivo* where DNA is likely to be bent (3, 4).

Since small internal forces are sufficient to increase binding affinity, and larger perturbations have minimal further effect (Fig. 4B), it appears that architectural *strain* or bending induced by small amounts of internal force is sensed by IHF. This characteristic may partially explain IHF's non-specific binding *in vivo*. Even if a DNA sequence does not conform to IHF's site specific consensus sequence, small internal forces can substantially increase the affinity of IHF for that site. The free IHF concentration *in vivo* has been estimated to be ~15-35 nM (41), or a thousand-fold less than the total IHF concentration of 6-30 μM (50). Given that there are at most a few hundred specific IHF sites (17), a significant fraction of IHF is likely nonspecifically sequestered on genomic DNA (41). Perhaps the large reservoir of non-specific binding sites with low *in vitro* affinities can be partially explained by affinity rescue due to DNA architecture.

**Materials and Methods**

**Protein Purification**

IHF (plasmid - a gift of Phoebe Rice) was expressed in *E coli* following standard protocols. Following lysis, IHF in the supernatant of a 50% ammonium sulfate cut was precipitated with an 80% ammonium sulfate cut. The pellet was resuspended and dialyzed against 100mM NaCl, 25mM HEPES pH 7.0 and 10% glycerol. The resulting solution was loaded on a SP Sepharose column. Step washes were performed at 150mM and 250mM NaCl, and the IHF containing fraction was eluted at the 350mM NaCl step in 25mM HEPES pH 7.0, 0.1mM EDTA, and 5% glycerol. Purified IHF was concentrated using an Amicon Ultra-15 Centrifugal Filter, 5000 MWCO, and dialyzed against High Salt IHF Buffer: 220mM NaCl, 50mM Tris-HCl pH 7.4, 0.5mM EDTA, 5% glycerol. As judged by SDS gel, protein was 92 % pure. Protein concentration was quantitated using absorption at 276nm, with a $5800M^{-1}cm^{-1}$ extinction coefficient (28). The active fraction of protein was determined by gel shift assays performed under stoichiometric conditions (51). All concentrations of IHF are reported as active fraction concentrations.

**DNA Constructs**

Please see Supplementary Material for all DNA sequences.

Linear constructs were formed by annealing oligos with 5' terminal Cy3 or Cy5 dyes (Fig. 1A), or by annealing complements of various lengths to dually-labeled (Cy3 and Cy5) ssDNA oligos (Fig. 1B). Consistent with the predicted thermodynamic stability of the duplexes, native PAGE revealed that complements with lengths above 30 bases were nearly 100% annealed (data not shown). Dually-labeled ssDNA oligos were circularized with CircLigase ssDNA Ligase (Epicentre, CL4115K) (Fig. 1C), gel purified, and checked with exonuclease digests as previously described (1, 2). Annealing to complementary oligos was performed in 1X TBE buffer supplemented with 50 mM NaCl. Annealing mixtures were cooled from $80^{o}C$ degrees to $4^{o}C$ over 90 minutes and then stored at $4^{o}C$.

**Solution FRET Measurements for Binding Affinity**

Dually-labeled DNA was titrated with a concentrated solution of IHF. Two replicates, as well as a mock control titration with equivalent volumes of buffer were performed for each construct. All experiments were performed in IHF high salt buffer (see Protein Purification) at room temperature.

Five minutes were allowed for equilibration after addition of concentrated IHF (longer equilibration times did not result in changes in the FRET ratio). Mixing was performed by pipetting. The initial volume of buffer was 700 uL, with an initial DNA concentration of 0.5 nM for all titrations except for the 57M2C35 construct which had a 1.0 nM initial concentration. After the last titration point, the solution had been diluted by ~ 20 %. DNA dilution was taken into account in the analysis.

Emission spectra were taken on a FluoroLog-3 fluorimeter (Horiba Jobin Yvon), with 5 nm excitation slit, 15 nm emission slit, 0.5 s integration time, and 2 nm increments. To maximize the signal to noise ratio, we elected not to place polarizers in the beam path. Emission spectra were recorded from 562-700 nm (542 nm Cy3 excitation) and 660-700 (640 nm direct Cy5 excitation). Cuvettes were cleaned with 5M Nitric Acid for at least 1 hour between experiments, and rinsed well with clean water.

The FRET ratio is defined as the ratio of acceptor emission resulting from FRET (Cy3 excitation), to acceptor emission when directly excited (Cy5 excitation): FRET ratio = F(670,542)/F(670,640), where F(ν,ν') refers to fluorescence intensity measured at wavelength ν (in nm) when excited at wavelength ν'. Prior to calculating F(670,542), the Cy3 emission tail at 670 nm was subtracted from the emission spectrum by fitting to a standard Cy3-labeled DNA emission spectrum (1, 47). The dissociation constant $K_D$ was derived by fitting FRET data to the following equation with SigmaPlot (v.10, Systat Software Inc.):

$$FRETRatio = (FractionBound)(\eta_b - \eta_f) + \eta_f$$

$$FractionBound = \frac{[PD]}{[D_T]}$$

$$[PD]^2 - (P_T + D_T + K_D)[PD] + P_T D_T = 0$$

where $\eta_b$ and $\eta_f$ are the FRET ratio corresponding to bound and free DNA respectively, $[P_T]$ and $[D_T]$ are the total IHF and DNA concentrations respectively, $[PD]$ is the concentration of IHF-DNA bound complex, and $K_D$ is the dissociation constant.

## Saturation Binding Electrophoretic Mobility Shift Assays

Electrophoretic Mobility Shift Assays (EMSA) were performed as an independent method to obtain binding affinities. In these experiments, the IHF concentration was held constant at 21 nM while the concentration of dually-labeled DNA was varied from 2.5 to 50 nM in a 20 μL reaction volume. Samples were incubated for 15 - 30 minutes at room temperature in high salt IHF buffer (see Protein Purification). Increasing the incubation time did not affect the binding affinities. Samples were loaded onto 4-20% Tris-HCl Gradient gels (Bio-Rad, 161-1159), and electrophoresed in 1X TBE on ice at 13mA for 135 minutes that had been pre-run for 15 minutes. Cy5 fluorescence was observed on a Typhoon gel scanner (GE Healthcare, Model 8600) at a 500V PMT setting. ImageQuant (Molecular Dynamics) was used to quantitate the bound and free DNA. Four replicates were performed for all constructs except for 57M2C35 where three replicates were performed.

At the concentrations of DNA and IHF employed in our experiments, a significant fraction of the DNA was depleted by binding to IHF. We fit to a model that accounts for ligand depletion and nonspecific binding by using the methodology developed by Swillens (45). Binding curves were fit in SigmaPlot.

## Competition Electrophoretic Mobility Shift Assays for the Linear Mutant DNA

The binding affinity of IHF to the linear 57mer mutant (57M1L) construct was sufficiently low to warrant competition experiments for determination of dissociation constants. In these experiments, the amount of IHF and labeled probe is held constant, while the unlabeled competitor concentration is varied. The amount of IHF in the assays was fixed at 63 nM. For our labeled probe we used 57L35 at a fixed concentration of 5 nM. For our competitor, we used a fully double-stranded sequence containing the M1 mutant site, 57M1. Since high concentrations of competitor were used (up to 50μM), a fully double stranded competitor was necessary to ensure no nonspecific binding to single stranded regions of the competitor.

IHF, buffer and labeled probe were combined and incubated for five minutes. Increasing competitor concentrations (0 µM to 50 µM) were added to compete the IHF off the labeled probe. The 20 µL reactions were allowed to equilibrate for 90 minutes, in the dark at room temperature. Samples were loaded onto a 10% TBE gel, pre-run at 13 mA for 15 minutes on ice, and subsequently run at 13 mA for 45 minutes on ice. Free and bound labeled probe were quantitated as in the saturation binding gels. Four replicates were performed.

The fraction of labeled probe bound to the protein was determined from the results of the quantification. We plotted the relative yield of the labeled probe-protein complex, *f*, against competitor concentration $[C]$ (41) with an offset to account for non-specific binding, $\alpha$. We fit the results using SigmaPlot according to the equation:

$$f = \frac{IC_{50}}{[C] + IC_{50}} + \alpha \qquad (1)$$

This yielded an estimate for $IC_{50}$ (the concentration of competitor necessary to compete IHF off half of the labeled DNA), which was used to calculate the $K_D$ for the unlabeled competitor. Since the concentrations of IHF exceeded the $K_D$s for both labeled and unlabeled species, we used the methodology of Linden (1982) (52), which allows calculation of $K_D$ without constraints on the IHF concentration. First, using the $IC_{50}$ from Equation 1, the free, unlabeled competitor concentration $I_F$ was determined from

$$I_F = IC_{50} - R_T + \frac{R_T}{2}\left[\frac{H_F}{K_{D,57L35} + H_F} + \frac{K_{D,57L35}}{K_{D,57L35} + H_F + \frac{R_T}{2}}\right] \qquad (2)$$

where $R_T$ is the total IHF concentration, $H_F$ is the concentration of free, labeled probe and $K_{D,57L35}$ is the dissociation constant of the labeled 57L35. Finally, $K_{D,M}$, the dissociation constant of the unlabeled mutant was determined from the quantity

$$K_{D,M} = \frac{I_F}{1 + \frac{H_F}{K_{D,57L35}} + \frac{R_T}{K_{D,57L35}}\left[\frac{K_{D,57L35} + \frac{H_F}{2}}{K_{D,57L35} + H_F}\right]} \qquad (3)$$

Results for the M2 mutant were highly variable, likely due to the large amount of added competitor necessitated by the use of fluorescence as opposed to radioactivity. Hence in this paper, we use the previously-measured value of 690 nM (41) for the $K_D$ of 57M2L.

**Bulk FRET Efficiencies for Internal Force Estimation**

FRET efficiencies were extracted from bulk fluorescence measurements as previously described (1, 2). Bulk fluorimetry was performed in either high salt IHF buffer (see protein purification) or low salt buffer (1X TBE, 50mM NaCl). There was no significant difference between FRET measurements for a given construct performed in low salt buffer or in the original calibration buffer (10mM Tris pH 8.0, 50mM NaCl) (data not shown). After correcting for incomplete annealing and incomplete donor labeling, FRET efficiencies were converted to a common Förster distance basis to account for differences in local environment and converted to forces using the previously measured calibration curve (1, 2).

**Acknowledgements**

   We thank Harish Agarwal for code to analyze fluorimetry data, Alan Lowe for experimental suggestions, feedback and critical reading of the manuscript, Markus Seeliger for technical guidance, and David Wemmer for feedback and helpful discussions, and the Marqusee and Bustamante labs for the use of their equipment. MS acknowledges support from the Howard Hughes Medical Institute Predoctoral fellowship. HS thanks the Fannie and John Hertz foundation. DS acknowledges support from DOD-NDSEG and NSF predoctoral fellowships. This work was supported by the University of California, Berkeley (J.L.), the Hellman Faculty Fund (J.L.), the Sloan and Searle foundations (J.L.), and the Department of Energy Office of Science, Energy Biosciences Program (J.L.).

## References

1. Shroff, H., D. Sivak, J. Siegel, A. McEvoy, M. Siu, A. Spakowitz, P. Geissler, and J. Liphardt. 2007. Optical Measurement of Mechanical Forces Inside Short DNA Loops. Biophys J in press.

2. Shroff, H., B. M. Reinhard, M. Siu, H. Agarwal, A. Spakowitz, and J. Liphardt. 2005. Biocompatible force sensor with optical readout and dimensions of 6 nm$^3$. Nano Lett. 5:1509-1514.

3. PerezMartin, J., and V. deLorenzo. 1997. Clues and consequences of DNA bending in transcription. Annual Review of Microbiology 51:593-628.

4. Schleif, R. 1992. DNA Looping. Annual Review of Biochemistry 61:199-223.

5. Landy, A. 1989. Dynamic, Structural, and Regulatory Aspects of Lambda-Site-Specific Recombination. Annual Review of Biochemistry 58:913-949.

6. Travers, A. A., and A. Klug. 1987. The Bending of DNA in Nucleosomes and Its Wider Implications. Philosophical Transactions of the Royal Society of London Series B-Biological Sciences 317:537-561.

7. Schultz, S. C., G. C. Shields, and T. A. Steitz. 1991. Crystal-Structure of a Cap-DNA Complex - the DNA Is Bent by 90-Degrees. Science 253:1001-1007.

8. Kim, Y. C., J. H. Geiger, S. Hahn, and P. B. Sigler. 1993. Crystal Structure of a Yeast Tbp Tata-Box Complex. Nature 365:512-520.

9. Kim, J. L., D. B. Nikolov, and S. K. Burley. 1993. Co-Crystal Structure of Tbp Recognizing the Minor-Groove of a Tata Element. Nature 365:520-527.

10. Brennan, R., S. Roderick, Y. Takeda, and B. Matthews. 1990. Protein-DNA Conformational Changes in the Crystal Structure of a lambda Cro-Operator Complex. PNAS 87:8165-8169.

11. Rice, P. A., S. W. Yang, K. Mizuuchi, and H. A. NAsh. 1996. Crystal structure of an IHF-DNA complex: A protein-induced DNA u-turn. Cell 87:1295-1306.

12. Wu, H. M., and D. M. Crothers. 1984. The Locus of Sequence-Directed and Protein-Induced DNA Bending. Nature 308:509-513.

13. Sun, D., L. H. Hurley, and R. M. Harshey. 1996. Structural distortions induced by integration host factor (IHF) at the H' site of phage lambda probed by (+)-CC-1065, pluramycin, and KMnO4 and by DNA cyclization studies. Biochemistry 35:10815-10827.

14. Teter, B., S. D. Goodman, and D. J. Galas. 2000. DNA bending and twisting properties of integration host factor determined by DNA cyclization. Plasmid 43:73-84.

15. Kahn, J. D., and D. M. Crothers. 1992. Protein-Induced Bending and DNA Cyclization. Proceedings of the National Academy of Sciences of the United States of America 89:6343-6347.

16. Parvin, J. D., R. J. McCormick, P. A. Sharp, and D. E. Fisher. 1995. Pre-Bending of a Promoter Sequence Enhances Affinity for the Tata-Binding Factor. Nature 373:724-727.

17. Nash, H. A. 1996. The HU and IHF Proteins: Accessory Factors for Complex Protein-DNA Assemblies. In Regulation of Gene Expression in Escherichia coli. Lin, and Lynch, editors. R. G. Landes Company. 149-179.

18. Sandman, K., S. L. Pereira, and J. N. Reeve. 1998. Diversity of prokaryotic chromosomal proteins and the origin of the nucleosome. Cellular and Molecular Life Sciences 54:1350-1364.

19. Holbrook, J. A., O. V. Tsodikov, R. M. Saecker, and M. T. Record. 2001. Specific and non-specific interactions of integration host factor with DNA: Thermodynamic evidence for disruption of multiple IHF surface salt-bridges coupled to DNA binding. Journal of Molecular Biology 310:379-401.

20. Barnard, A., A. Wolfe, and S. Busby. 2004. Regulation at complex bacterial promoters: how bacteria use different promoter organizations to produce different regulatory outcomes. Current Opinion in Microbiology 7:102-108.

21. Buck, M., M. T. Gallegos, D. J. Studholme, Y. L. Guo, and J. D. Gralla. 2000. The bacterial enhancer-dependent sigma(54) (sigma(N)) transcription factor. Journal of Bacteriology 182:4129-4136.

22. Goosen, N., and P. Vandeputte. 1995. The Regulation of Transcription Initiation by Integration Host Factor. Molecular Microbiology 16:1-7.

23. Craig, N. L., and H. A. Nash. 1984. Escherichia-Coli Integration Host Factor Binds to Specific Sites in DNA. Cell 39:707-716.

24. Nash, H. A., and C. A. Robertson. 1981. Purification and Properties of the Escherichia-Coli Protein Factor Required for Gamma-Integrative Recombination. Journal of Biological Chemistry 256:9246-9253.

25. Thompson, J. F., and A. Landy. 1988. Empirical Estimation of Protein-Induced DNA Bending Angles - Applications to Lambda-Site-Specific Recombination Complexes. Nucleic Acids Research 16:9687-9705.

26. Lorenz, M., A. Hillisch, S. D. Goodman, and S. Diekmann. 1999. Global structure similarities of intact and nicked DNA complexed with IHF measured in solution by fluorescence resonance energy transfer. Nucleic Acids Res 27:4619-4625.

27. Rice, P. A. 1997. Making DNA do a U-turn: IHF and related proteins. Curr Opin Struct Biol 7:86-93.

28. Lynch, T. W., E. K. Read, A. N. Mattis, J. F. Gardner, and P. A. Rice. 2003. Integration host factor: putting a twist on protein-DNA recognition. J Mol Biol 330:493-502.

29. Grove, A., A. Galeone, L. Mayol, and E. P. Geiduschek. 1996. Localized DNA flexibility contributes to target site selection by DNA-bending proteins. J Mol Biol 260:120-125.

30. Zahn, K., and F. R. Blattner. 1987. Direct Evidence for DNA Bending at the Lambda Replication Origin. Science 236:416-422.

31. Bustamante, C., S. B. Smith, J. Liphardt, and D. Smith. 2000. Single-molecule studies of DNA mechanics. Curr Opin Struc Bio 10:279-285.

32. Murphy, M. C., I. Rasnik, W. Cheng, T. M. Lohman, and T. J. Ha. 2004. Probing single-stranded DNA conformational flexibility using fluorescence spectroscopy. Biophys J 86:2530-2537.

33. Du, Q., M. Vologodskaia, H. Kuhn, M. Frank-Kamenetskii, and A. Vologodskii. 2005. Gapped DNA and cyclization of short DNA fragments. Biophys J 88:4137-4145.

34. Cloutier, T. E., and J. Widom. 2004. Spontaneous sharp bending of double-stranded DNA. Mol Cell 14:355-362.

35. Seol, Y., J. Li, P. C. Nelson, T. T. Perkins, and M. D. Betterton. 2007. Elasticity of short DNA molecules: theory and experiment for contour lengths of 0.6-7 microm. Biophys J 93:4360-4373.

36. Kuznetsov, S. V., S. Sugimura, P. Vivas, D. M. Crothers, and A. Ansari. 2006. Direct observation of DNA bending/unbending kinetics in complex with DNA-bending protein IHF. PNAS 103:18515-18520.

37. Lorenz, M., and S. Diekmann. 2001. Quantitative distance information on protein-DNA complexes determined in polyacrylamide gels by fluorescence resonance energy transfer. Electrophoresis 22:990-998.

38. Sugimura, S., and D. M. Crothers. 2006. Stepwise binding and bending of DNA by Escherichia coli integration host factor. PNAS 103:18510-18514.

39. Fried, M., and D. M. Crothers. 1981. Equilibria and kinetics of lac repressor-operator interactions by polyacrylamide gel electrophoresis. Nucleic Acids Res 9:6505-6525.

40. Garner, M. M., and A. Revzin. 1981. A gel electrophoresis method for quantifying the binding of proteins to specific DNA regions: application to components of the Escherichia coli lactose operon regulatory system. Nucleic Acids Res 9:3047-3060.

41. Yang, S. W., and H. A. Nash. 1995. Comparison of protein binding to DNA in vivo and in vitro: Defining an effective intracellular target. EMBO 14:6292-6300.

42. Wang, S., R. Cosstick, J. F. Gardner, and R. I. Gumport. 1995. The specific binding of Escherichia coli integration host factor involves both major and minor grooves of DNA. Biochemistry-Us 34:13082-13090.

43. Murtin, C., M. Engelhorn, J. Geiselmann, and F. Boccard. 1998. A quantitative UV laser footprinting analysis of the interaction of IHF with specific binding sites: re-evaluation of the effective concentration of IHF in the cell. J Mol Biol 284:949-961.

44. Dhavan, G. M., D. M. Crothers, M. R. Chance, and M. Brenowitz. 2002. Concerted binding and bending of DNA by Escherichia coli integration host factor. J Mol Bio 315:1027-1037.

45. Swillens, S. 1995. Interpretation of Binding Curves Obtained with High Receptor Concentrations - Practical Aid for Computer-Analysis. Molecular pharmacology 47:1197-1203.

46. Yang, C. C., and H. A. Nash. 1989. The interaction of E. coli IHF protein with its specific binding sites. Cell 57:869-880.

47. Clegg, R. M. 1992. Fluorescence Resonance Energy-Transfer and Nucleic-Acids. Methods in Enzymology 211:353-388.

48. Baumann, C. G., S. B. Smith, V. A. Bloomfield, and C. Bustamante. 1997. Ionic effects on the elasticity of single DNA molecules. PNAS 94:6185-6190.

49. Wenner, J. R., M. C. Williams, I. Rouzina, and V. A. Bloomfield. 2002. Salt dependence of the elasticity and overstretching transition of single DNA molecules. Biophys J 82:3160-3169.

50. Ditto, M. D., D. Roberts, and R. A. Weisberg. 1994. Growth phase variation of integration host factor level in Escherichia coli. J Bacteriol 176:3738-3748.

51. Hudson, J. M., and M. G. Fried. 1991. The binding of cyclic AMP receptor protein to two lactose promoter sites is not cooperative in vitro. J Bacteriol 173:59-66.

52. Linden, J. 1982. Calculating the Dissociation-Constant of an Unlabeled Compound from the Concentration Required to Displace Radiolabel Binding by 50-Percent. Journal of Cyclic Nucleotide Research 8:163-172.

**Table 1.** Dissociation constants for linear and circular wildtype constructs with varying internal forces as determined by solution FRET and/or EMSA.

| Dissociation constants for wildtype sequences, $K_D$ (nM) | | | | |
|---|---|---|---|---|
| Construct | Linear | | Circular | |
| | Solution FRET (mean ± stdev) n = 2 | EMSA (mean ± sterr) n ≥ 4 | Solution FRET (mean ± stdev) n = 2 | EMSA (mean ± sterr) n ≥ 4 |
| 57-30 | | 4.7 ± 1.4 | 0.85 ± 0.07 | 0.5 ±0.3 |
| 57-35 | | 6.8 ± 2.3 | 0.49 ± 0.01 | 0.2 ±0.1 |
| 57-39 | | 13.4± 8.8 | 0.54 ± 0.09 | 0.7 ± 0.4 |
| 57-43 | | 6.5 ± 4.4 | 0.80 ± 0.02 | 0.2 ± 0.1 |
| | | | | |
| 43-41 | 9.32 ± 0.16 | | | |
| | | | | |
| 101-77 | | 1.5 ± 0.3 | | 0.9 ± 0.2 |
| 101-87 | | 5.1 ± 1.7 | | 0.5 ± 0.2 |

**Table 2.** Dissociation constants for linear and circular mutant constructs as determined by solution FRET and/or EMSA.

| Dissociation constants for mutant sequences, $K_D$ (nM) | | | | |
|---|---|---|---|---|
| Construct | Linear | | Circular | |
| | Solution FRET | EMSA | Solution FRET | EMSA |
| | (mean ± stdev) | (mean ± sterr) | (mean ± stdev) | (mean ± sterr) |
| | n = 2 | n ≥ 4 | n = 2 | n ≥ 4[*] |
| 57M1-35 | | 46 ± 6 | 0.89 ± 0.06 | 0.6 ± 0.3 |
| 57M2-35 | | 690 ± 69[†] | 4.67 ± 1.78 | 1.6 ± 0.5[*] |
| | | | | |
| 43M1-41 | 156 ± 6 | | | |

[*] 3 replicates were performed for 57M2C35
[†] from (41) where errors were estimated as 10% of the measured $K_D$

**Figure Legends**

**Figure 1.**
**Schematic of DNA constructs.** Two types of linear constructs were used. The first type (**A**) is formed by annealing complementary ssDNA oligos, each labeled with either Cy3 (green circle) or Cy5 (red circle) dyes on their 5' termini. The second type of linear construct (**B**) is formed by annealing dually-labeled ssDNA (blue) directly to unlabeled complementary ssDNA (red) of various lengths. Circular constructs (**C**) were formed by first incubating ssDNA, dually-labeled with Cy3 and Cy5 dyes, with ssDNA ligase. The Cy3 and Cy5 FRET pair separated by 10 bases of ssDNA forms the optical force sensor. Gel purified ssDNA circles were then annealed to complements of different lengths. All constructs contained a dsDNA region with a wildtype λ H' IHF binding site or a mutant binding site. IHF (brown ellipse) induces a dramatic bend in the DNA.

**Figure 2.**
**Solution FRET assay for IHF binding affinity to wildtype DNA constructs with varying internal forces.** Binding curves were measured by fluorimetry of dually labeled DNA (0.5 nM) titrated with concentrated IHF. (**A & B**) Representative raw emission spectra with Cy3 excitation (542nm), and Cy5 excitation (inset - 640nm), taken at the beginning (black, - IHF) and endpoint of the IHF titration (red, + IHF, 146 nM for linear 43L41, 6 nM for circle 57C35). The overall decrease in fluorescence intensity at the end of the titration is due to dilution and bleaching, but these effects are accounted for by the ratiometric FRET measure (see Methods) used to calculate fraction DNA bound. Normalized FRET ratio indicates the fraction of DNA bound for the linear (**C**) and circular (**D**) titrations. Circular constructs with four different dsDNA:ssDNA ratios were studied. Two replicates (n = 2) were performed for all constructs. Filled symbols are IHF titrations, lines are fits, and open circles are buffer controls.

**Figure 3.**
**Electrophoretic Mobility Shift Assay (EMSA) for IHF binding affinity to wildtype DNA constructs with varying internal forces.** A fixed amount of IHF (21 nM) was titrated with increasing amounts (shown) of linear or circular DNA. Dye-labeled DNA experienced reduced mobility for both linear (57L35 - A) and circular (57C35, B) constructs. Four replicates (n = 4) were performed for each construct. Free and bound bands as assayed by Cy5 fluorescence were quantitated by densiometry to obtain binding curves, shown with associated fits. A total of 59 EMSA assays were performed for the different constructs.

**Figure 4.**
**Internal force above ~3 pN does not significantly affect IHF binding affinity. A**) $K_D$ vs. FRET efficiencies, with dissociation constants from EMSA (see **Fig. 3**), and FRET efficiencies from optical force sensors (**Fig. S2** (1)). **B**) $K_D$ vs. Internal Force Estimate. Internal force is estimated by using a previously determined calibration curve (2) to map the converted FRET efficiencies to force. We do not account for the higher ionic strength used in the EMSA experiments as the expected force differences are below our experimental error (see text and **Table S1**). Linear constructs (**Table 1**) have zero internal force by definition. Black squares:

57mer data. Black empty square denotes 57C30 whose FRET efficiency falls above the range of the calibration curve. Red squares: 101mer data. Note the logarithmic y-axis.

**Figure 5.**
**Solution FRET assay for IHF binding affinity to mutant DNA constructs.** Binding curves for linear (**A**) and circular (**B**) wildtype and mutant constructs. Note the different scales on the x-axes. Black symbols are wildtype constructs, red symbols are mutant (M1) constructs, open circles are buffer control titrations, lines are corresponding fits. The linear wildtype titration (43L41) was only carried out until 146 nM IHF at which point a reasonable plateau had been reached. Inset in (**B**): binding curve to circular mutant (M2) construct. Red symbols are an IHF titration, green circles are buffer controls.

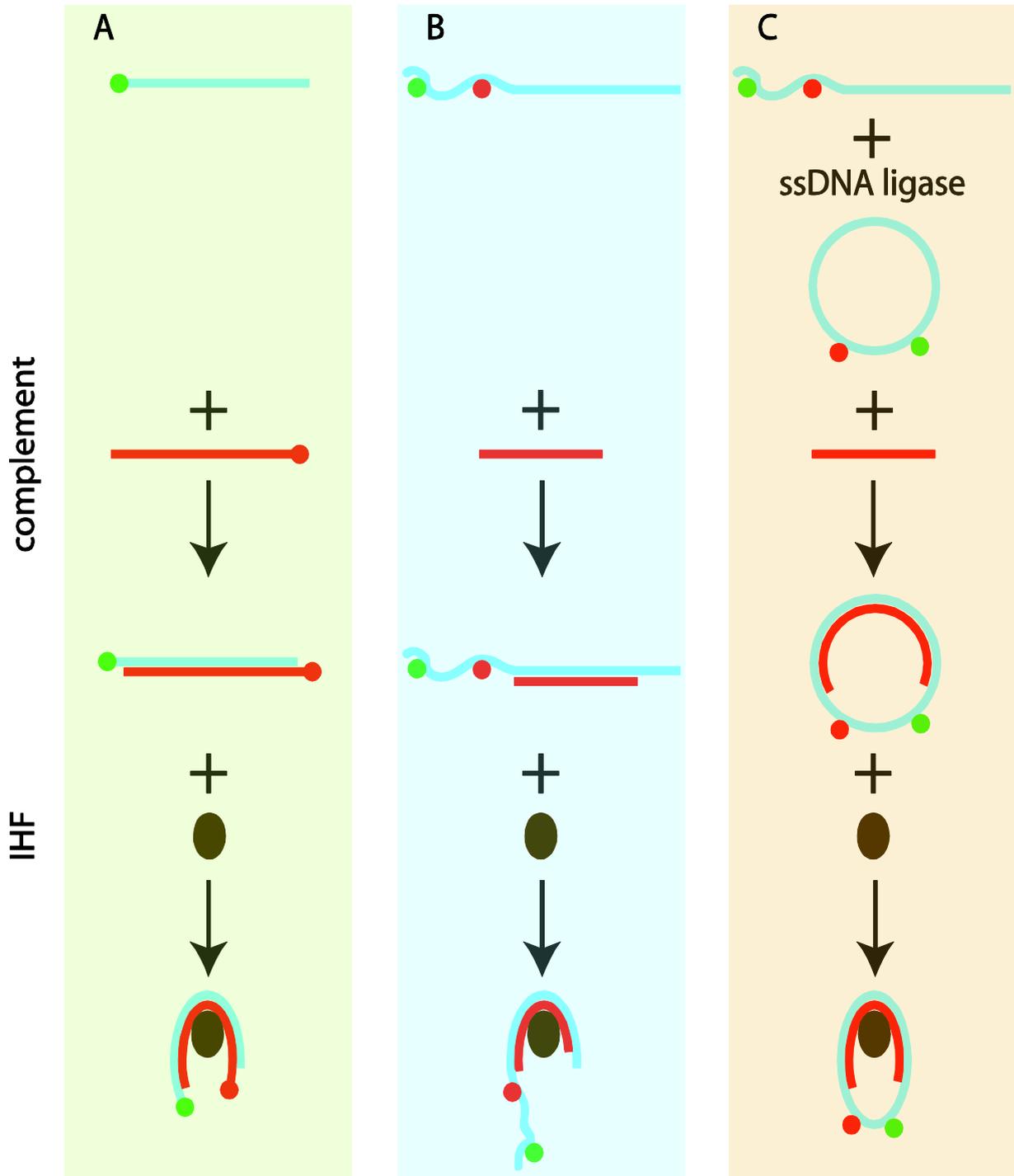

Figure 1:

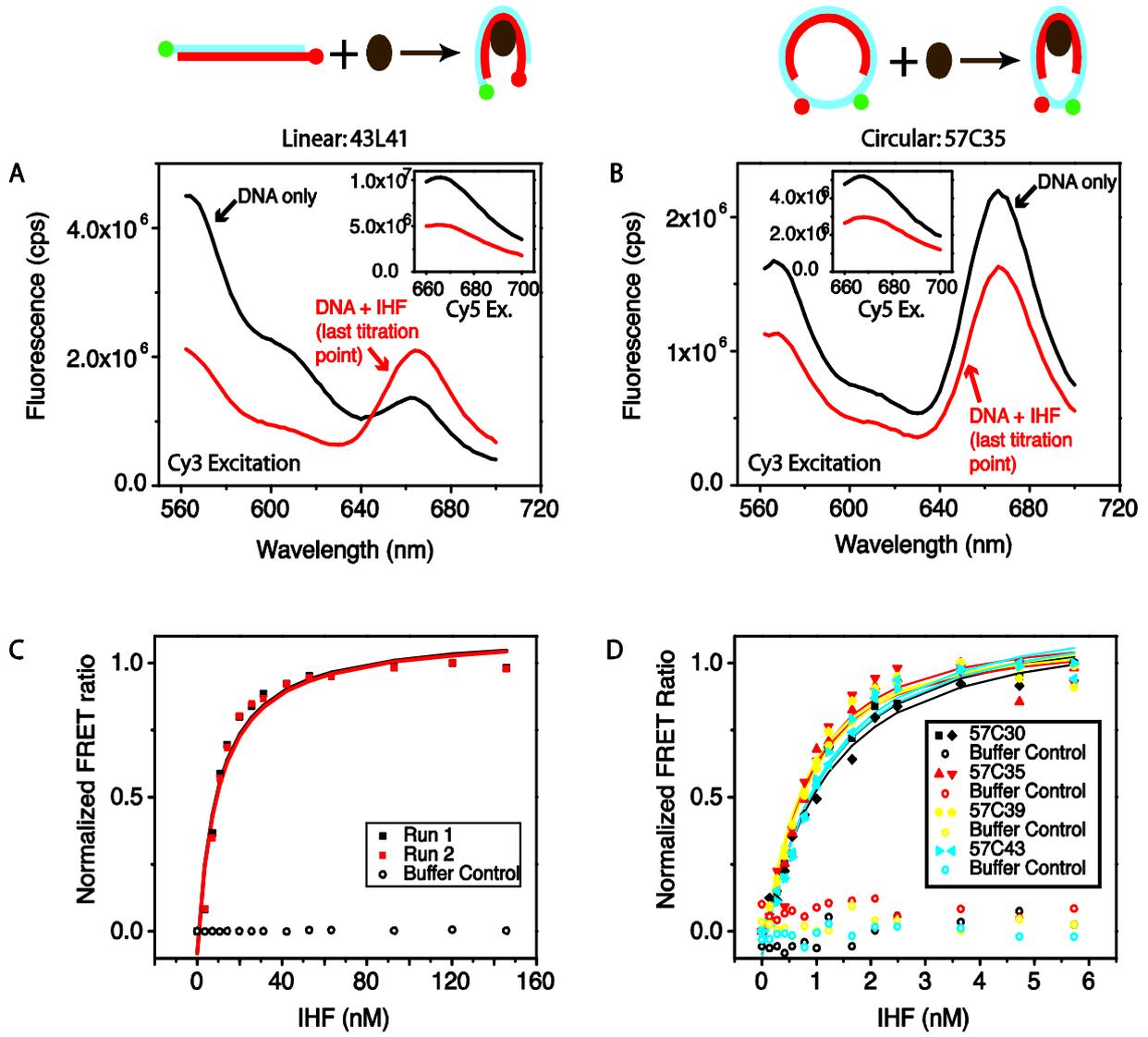

Figure 2:

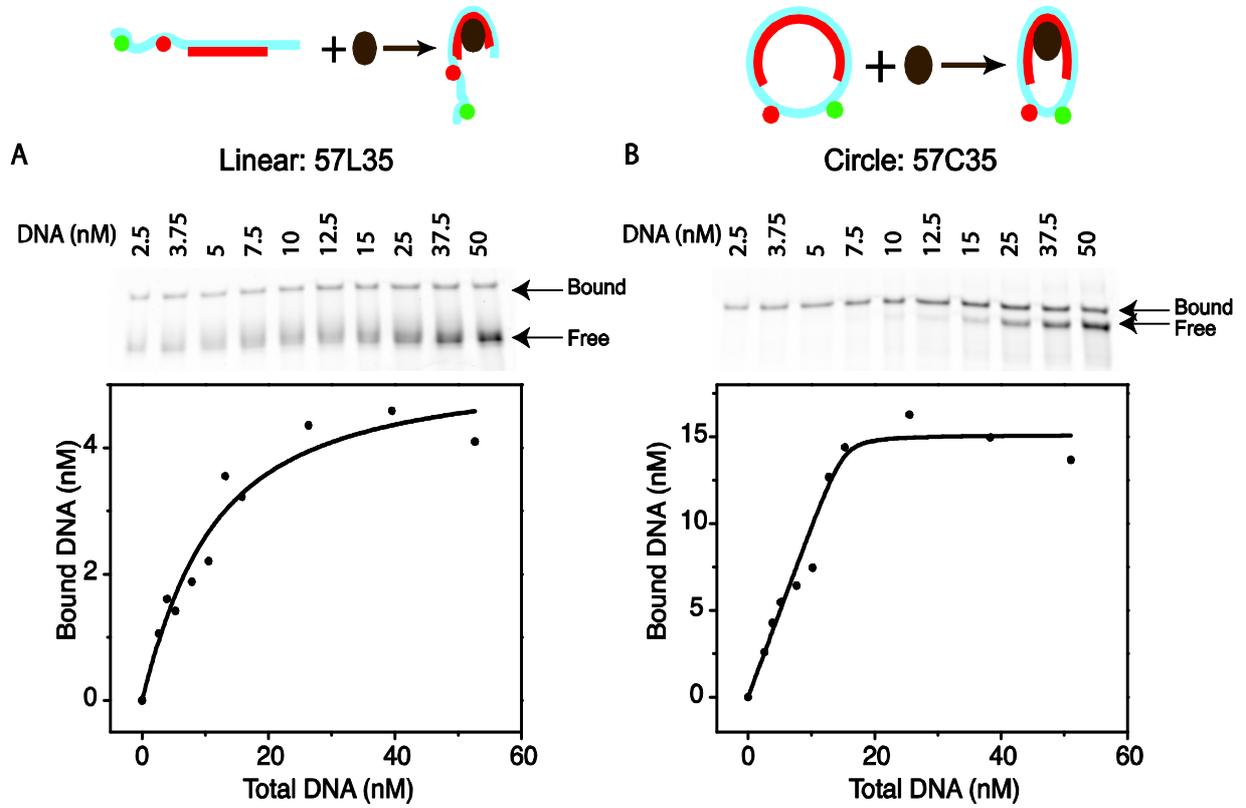

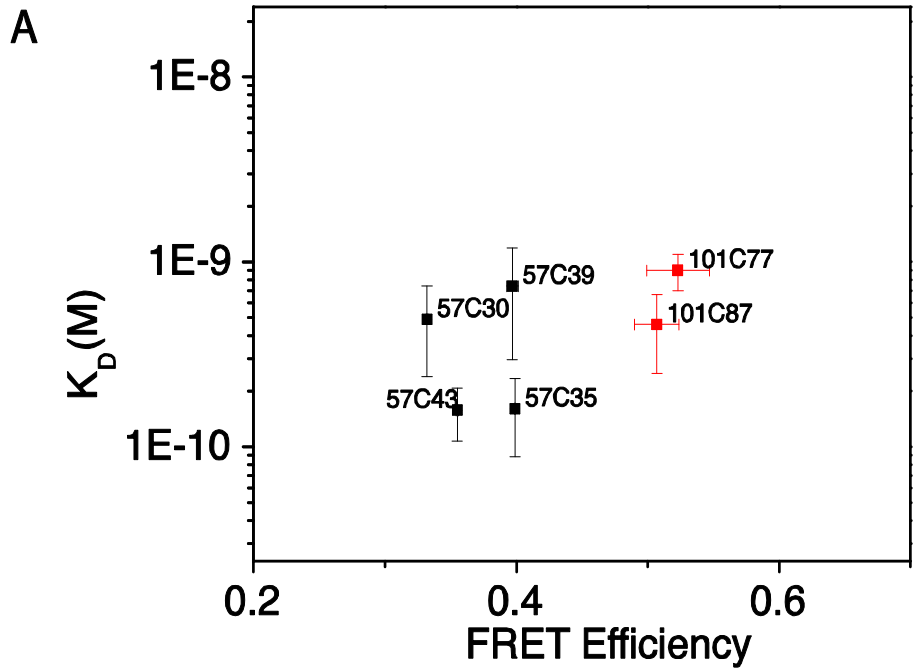

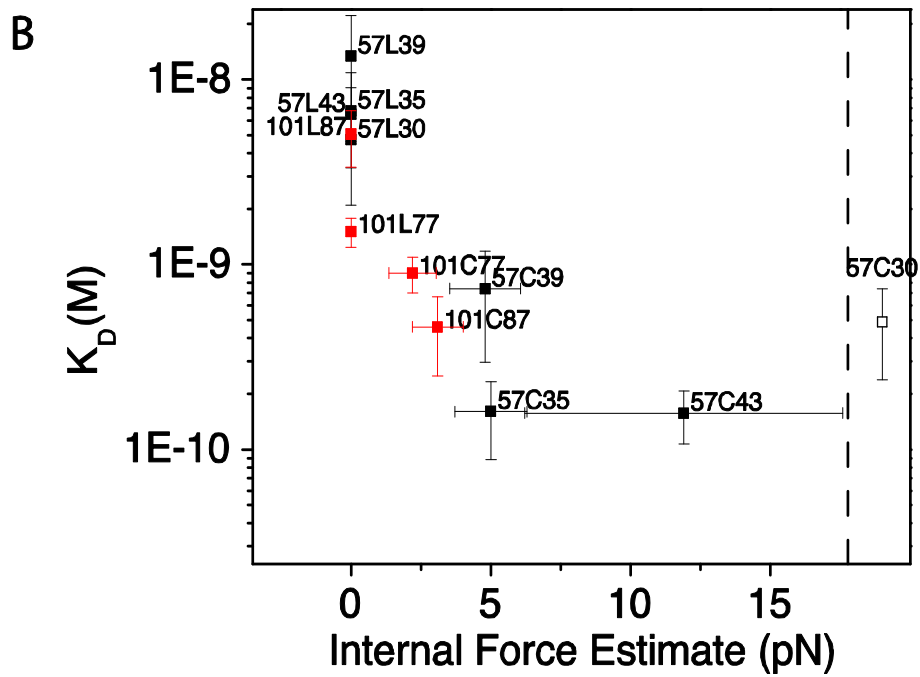

Figure 4:

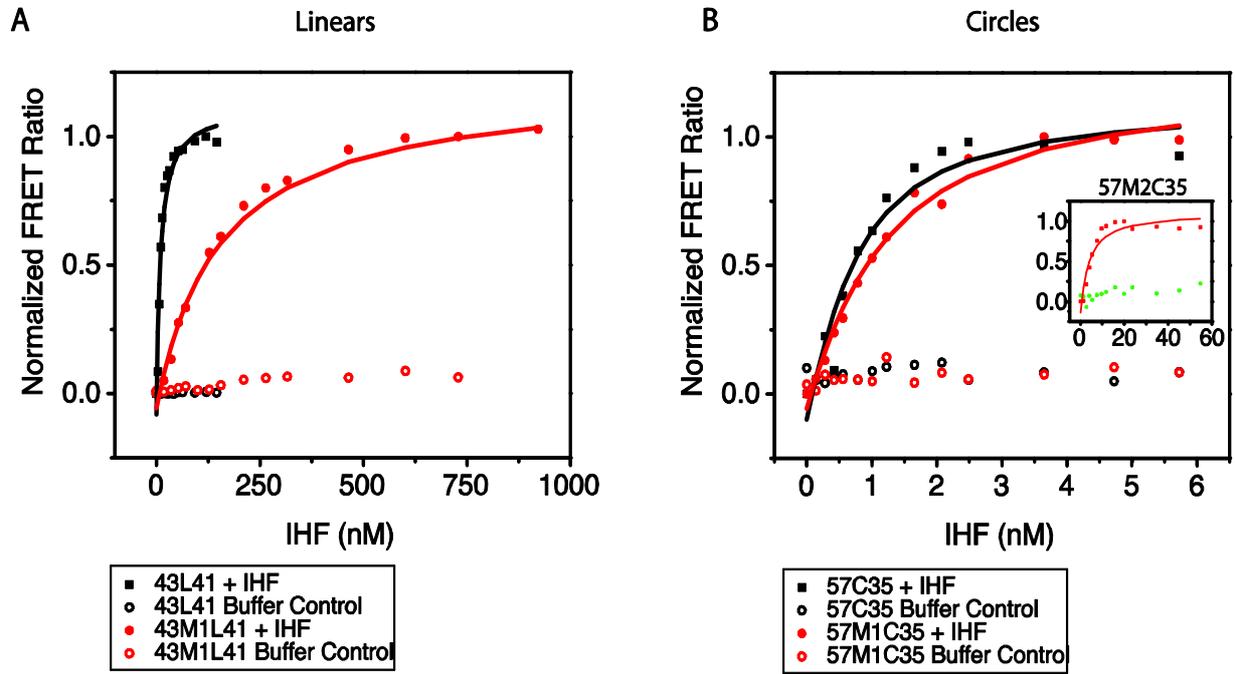

Figure 5:

**Supplementary Material for "Mechanical conversion of low-affinity Integration Host Factor binding sites into high-affinity sites"**

Merek Siu, Hari Shroff , Jake Siegel, Ann McEvoy, David Sivak, Ann Maris, Andrew Spakowitz, Jan Liphardt

**Oligonucleotide Sequences**

Linear Oligos for Solution FRET experiments (**Fig. 1A**)

**43Cy3** 5' Cy3-GGC CGG GCC AAA AAA GCA TTG CTT ATC AAT TTG TTG CAC CTC T
**43CompCy5**
5' Cy5-GTA GAG GTG CAA CAA ATT GAT AAG CAA TGC TTT TTT GGC CCG G

**43M1Cy3**
5' Cy3-GGCCGGGCCAAAAAAGCATTGCTTATCAATTTGTCGCACCTCT
**43M1CompCy5**
5' Cy5-GTAGAGGTGCGACAAATTGATAAGCAATGCTTTTTTGGCCCGG

Dually labeled oligos (**Fig. 1B&C**)

Cy3 and Cy5 dyes were covalently attached to single stranded DNA (Fidelity Systems Non-Vanilla Oligonucleotide Synthesis) to make templates for the loop constructs. **Bold** denotes the 35 bp IHF consensus sequence, <span style="color:red">red</span> highlights bases that are changed in the mutant sequences.

**57mer Loop**, 5'P - AC[U-Cy5]ACGGATGGGA[T-Cy3]GGCCG**GGCCAAAA AAGCATTGCTTATCAATTTGTTGCACC**TCT

**57M1**, 5'P - AC[U-Cy5]ACGGATGGGA[T-Cy3]GGCCG**GGCCAAAA AAGCATTGCTTATCAATTTGT<span style="color:red">C</span>GCACC**TCT

**57M2**, 5'P - AC[U-Cy5]ACGGATGGGA[T-Cy3]GGCCG**GGCCAAAA AAGCATTGCTTATC<span style="color:red">CC</span>TTTG<span style="color:red">GG</span>GCACC**TCT

**101mer Loop**, 5'P – AC[U-Cy5]ACGGATGGGA[T-Cy3]GACAAGACTAGGATTACACAACTGCAGGGCCAAAAAAGCATTGCTTATCAATT TGTTGCACCTCTAGACTATGATTTATACAAGTAG

<u>Unlabeled Complements</u>
For some experiments, the following oligos (Integrated DNA Technologies) were annealed to the above sequences:

**57_30**, 5' GTGCAACAAATTGATAAGCAATGCTTTTTT
**57_35**, 5' AGGTGCAACAAATTGATAAGCAATGCTTTTTTGGC
**57_39**, 5' AGAGGTGCAACAAATTGATAAGCAATGCTTTTTTGGCCC
**57_43**, 5' GTAGAGGTGCAACAAATTGATAAGCAATGCTTTTTTGGCCCGG
**57M135**, 5'AGGTGCGACAAATTGATAAGCAATGCTTTTTTGGC
**57M235**, 5' AGGTGCCCCAAAGGGATAAGCAATGCTTTTTTGGC
**101_77** 5' TTGTATAAATCATAGTCTAGAGGTGCAACAAATTGATA
AGCAATGCTTTTT T GGCCCTGCAG TTGTGTAATCCTAGT
**101_87** 5' TCTACTTGTATAAATCATAGTCTAGAGGTGCAACAAATTGATA
AGCAATGC TTTTTTGGCCCTGCAGTTGTGTAATCCTAGTCTTGT

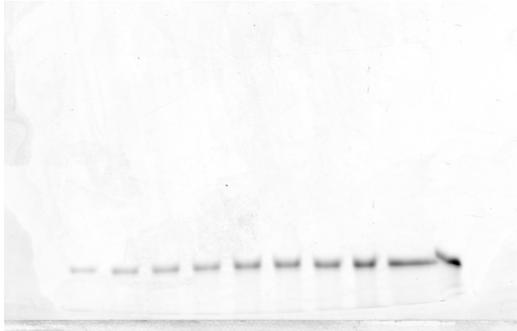

**Figure S1, IHF has no detectable binding to the single stranded force sensor**. Total concentrations of ssDNA and IHF in each lane were identical to total concentrations in other saturation binding experiments when full complements were used. Only one band, corresponding to free ssDNA, was observed at all concentrations of ssDNA.

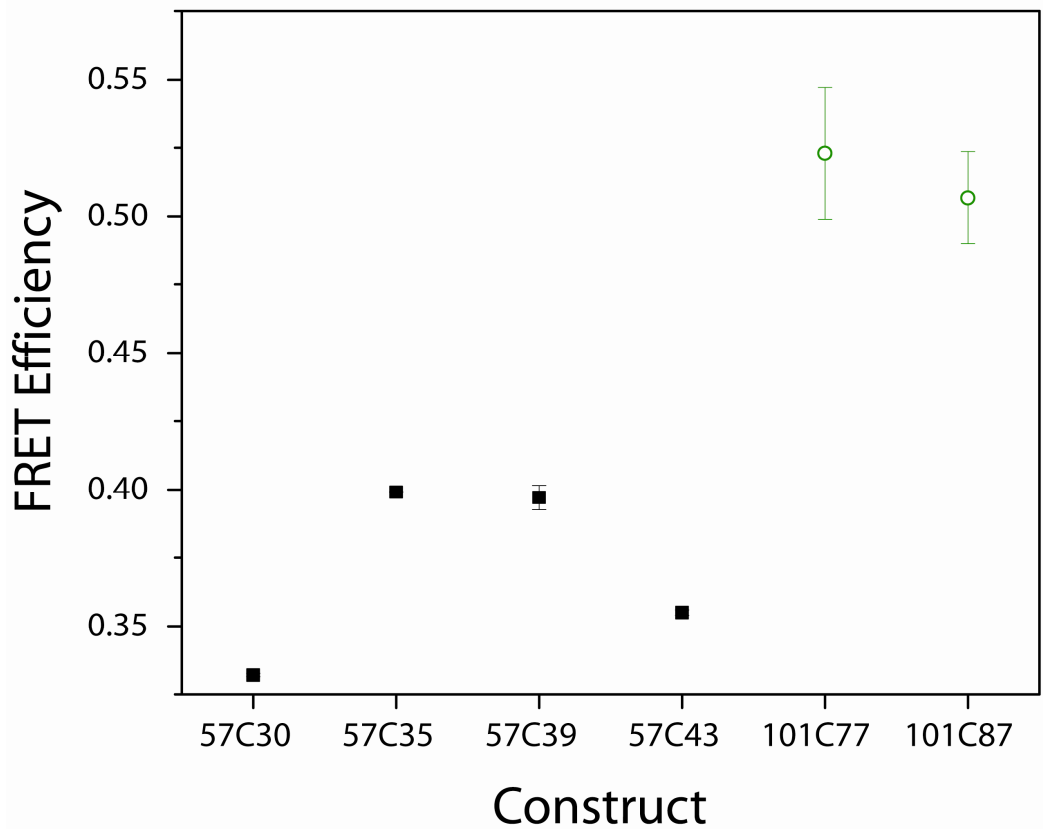

**Figure S2, 101mer constructs have higher FRET efficiencies than 57mer constructs (data from (1)).** FRET efficiencies for 57mer loops (black) with 30, 35, 39, 43 base double-stranded regions and 101mer loops (green circles) with 77 and 87 base double-stranded regions were measured in low ionic strength buffer. FRET efficiencies for the linear constructs are not shown since they are not related to the internal force, which is zero by definition.

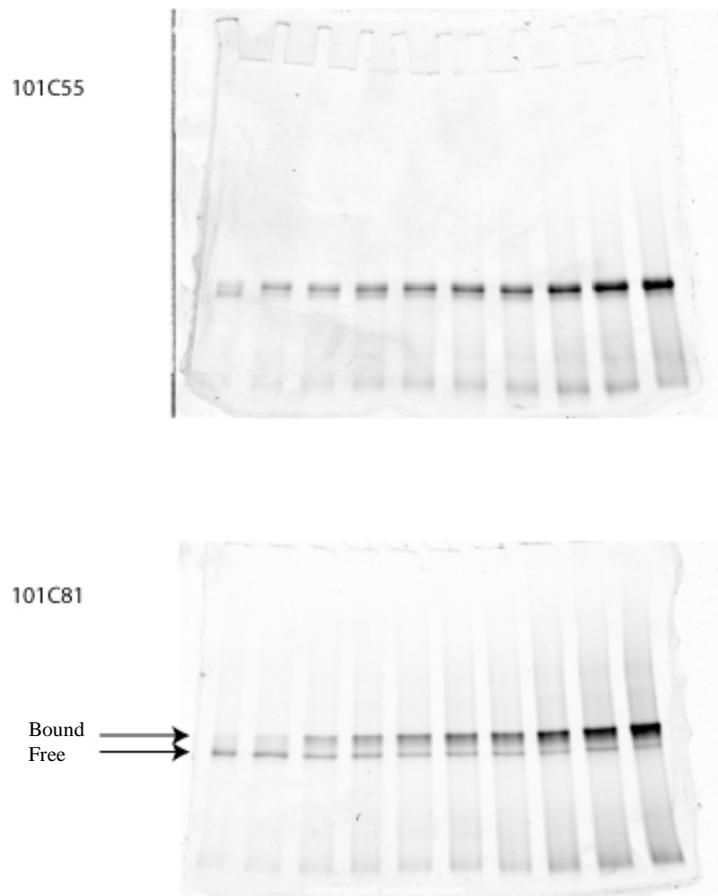

101C55

101C81

Bound
Free

**Figure S3, Limits to $K_D$ determination.** Representative gels for 101C55 and 101C81 are shown. For 101C55 constructs, free and bound bands are inseparable. Arrows indicate free and bound bands in the 101C81 gel.

Monte Carlo simulations of salt effects on internal force estimation

To estimate the error in the internal force conversion due to the elevated experimental salt concentrations, we performed Metropolis Monte Carlo simulations. Simulations were performed without any free parameters in high salt conditions used for our binding assays (220mM NaCl) and sensor calibration salt conditions (50mM NaCl). DNA was represented as a wormlike chain discretized at the level of a single nucleotide, with ssDNA nucleotides and dsDNA basepairs distinguished by different contour lengths and persistence lengths. Our Hamiltonian included DNA bending elasticity, twist elasticity, and volume interactions. We sampled thermal fluctuations using standard Metropolis Monte Carlo methods. Trial displacements preserving geometric constraints included 'crankshaft rotations' and twist rotations. The Förster distance, $R_0$, and single-stranded persistence length, $l_{Pss}$, parameters were fit through calibration to previous single-molecule force experiments (see (1) for details). Average forces were calculated directly from histograms of dsDNA helix end-to-end separation.

| | Simulated Forces (pN $\pm$ ~1 pN) | |
|---|---|---|
| | Low salt | High salt |
| 57C30 | 2 | 2 |
| 57C35 | 4 | 5 |
| 57C39 | 5 | 5 |
| 57C43 | 4 | 4 |
| 101C77 | 2 | 1 |
| 101C87 | 1 | 1 |

**Table S1, Monte Carlo simulations of the internal force in DNA constructs in low and high salt buffers.** EMSA experiments were performed in a high salt buffer whereas single molecule calibration of the force sensor was performed in a low salt buffer. Monte Carlo simulations (described above and in (1)) were performed using estimated salt-dependent changes in the persistence length of single-stranded and double-stranded DNA (2-4). This set of simulations has a precision of $\pm$ ~1 pN.

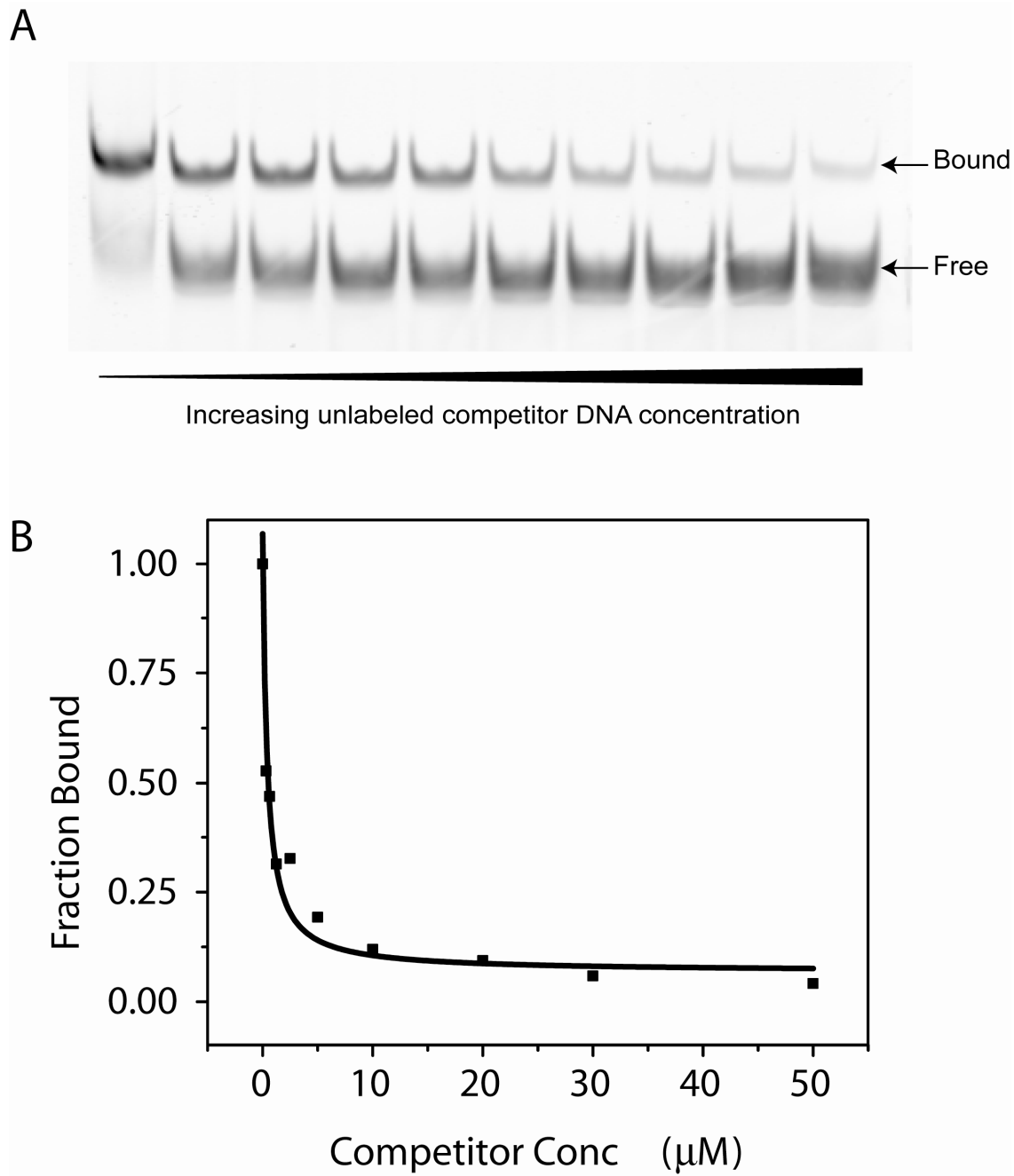

**Figure S4,** Competition EMSA to determine M1 mutant binding affinity.
IHF (63 nM) was incubated with labeled wildtype DNA (57L35) (5 nM) and competed off by
titrating with unlabeled competitor mutant (57M157) DNA. (A) Representative gel shows the
concentration of retarded bound labeled DNA decreasing as the unlabeled competitor DNA
concentration increases. (B) Quantitation of the gel showing fraction bound with increasing
competitor DNA concentration and associated fit.

**References for Supplementary Information**

1.  Shroff, H., D. Sivak, J. Siegel, A. McEvoy, M. Siu, A. Spakowitz, P. Geissler, and J. Liphardt. 2007. Optical Measurement of Mechanical Forces Inside Short DNA Loops. Biophys J in press.
2.  Murphy, M. C., I. Rasnik, W. Cheng, T. M. Lohman, and T. J. Ha. 2004. Probing single-stranded DNA conformational flexibility using fluorescence spectroscopy. Biophys J 86:2530-2537.
3.  Baumann, C. G., S. B. Smith, V. A. Bloomfield, and C. Bustamante. 1997. Ionic effects on the elasticity of single DNA molecules. PNAS 94:6185-6190.
4.  Wenner, J. R., M. C. Williams, I. Rouzina, and V. A. Bloomfield. 2002. Salt dependence of the elasticity and overstretching transition of single DNA molecules. Biophys J 82:3160-3169.